# Polarization-independent self-hybridized strong coupling by Brillouin zone folding induced quasi-bound states in the continuum in bulk $WS_2$ metasurfaces

Xiongwei Guo, Fu Liu, Jinhai Si, Xun Hou, Kuidong Wang*

*Key Laboratory for Physical Electronics and Devices of the Ministry of Education and Shaanxi Key Lab of Information Photonic Technique, School of Electronic Science and Engineering, Xi'an Jiaotong University, Xi'an, 710049, China*

*Contact author: wangkuidong@xjtu.edu.cn

**Keywords:** BZF-QBIC mode, polarization-independent, bulk $WS_2$, strong coupling

## Abstract

High-refractive-index dielectric metasurfaces supporting optical modes are ideal candidates for achieving strong coupling with transition-metal dichalcogenides (TMDCs) materials. However, optical illumination with a specific polarization is always required to estimate the maximal coupling strength in such coupled systems due to the polarization-dependent optical modes. Here, we propose a self-hybridized strong coupling between Brillouin zone folding induced quasi-bound states in the continuum (BZF-QBICs) mode and *A* excitons in the same bulk $WS_2$ metasurface. By changing the thickness of the bulk $WS_2$ metasurfaces, a typical anticrossing behavior with Rabi splitting of 165.9 meV is observed. Such Rabi splitting is unchanged when changing the polarization of the incident light from linearly to circularly polarizations, demonstrating the polarization-independent self-hybridized strong coupling. This work provides a novel method for realizing robust polarization-independent polaritons in bulk TMDCs metasurface and offers new insights for expanding the applications of polaritonic devices with arbitrary polarizations.

# 1. Introduction

Strong coupling between excitons of TMDCs and photons of optical cavities has emerged as an important platform for exploring new physical characteristics of low-dimensional materials, including Bose–Einstein condensation[1,2], superfluidity,[3] valley-polarized polaritons[4,5], and also paving the way for applications in polariton lasers[6,7], polaritonic photodiodes[8,9] and switchings[10,11]. These strongly coupled systems can form two branches of polaritons, i.e., upper and lower exciton-polaritons, by Rabi splitting when the energy exchange rate between excitons and photons exceeds each of their dissipation rates. Since strong light–matter coupling requires both large oscillator strength of excitons and high-quality (high-Q) optical modes[12,13], TMDCs are considered to be one of ideal van der Waals materials for realizing strong coupling due to their large binding energy[14,15] and large oscillation strength[16,17]. As for high-Q optical modes, Fabry–Perot (F–P) cavities[18,19] and plasmonic nanostructures[20,21] are typically employed to confine light within small mode volumes, thereby generating a high-intensity local electromagnetic field and enabling strong coupling with TMDCs materials. However, the conventional high-Q F-P cavities usually possess large footprint (inch size on transverse plane), while plasmonic structures suffer from strong intrinsic Ohmic losses at visible wavelengths, both of which hinder the applications of these TMDC-supported strong coupling on chip-scale polaritonic devices.

High-index dielectric metasurfaces supporting high-Q optical resonances provide efficient candidates for achieving TMDC-based strong coupling with small footprint[22-25]. These dielectric metasurfaces are usually compatible with the existing metal–oxide–semiconductor technology[26,27] and exhibit intrinsic low optical losses. As a typical high-Q optical mode supported by dielectric metasurfaces, photonic quasi-bound states in the continuum (QBIC) mode has especially attracted much attention for studying strong coupling[28-32], owing to its large localized field enhancement and flexible mode tunability[33]. Such QBIC mode comes from the mode coupling and destructive interference between several fundamental Mie resonances, thus suppressing the radiation losses[34]. However, the existing QBIC modes strongly rely on the breaking of the geometric symmetry of the designed structures, resulting in a polarization-selective optical property. As a result, the QBIC-based strong coupling in TMDCs is dependent on the polarization direction, which strongly restricts the realization of low-loss polaritonic devices with arbitrary polarizations in low-dimensional materials.

In this work, we propose a polarization-independent self-hybridized strong coupling in bulk $WS_2$ metasurfaces. This coupling comes from the interaction between *A* excitons of $WS_2$ and polarization-independent Brillouin zone folding induced QBIC (BZF-QBIC) modes of the structure of $WS_2$ itself. Compared to the conventional polarization-dependent optical modes in dielectric metasurfaces, the design of $C_{4v}$ symmetric bulk $WS_2$ metasurfaces enabling the polarization-independent intrinsic property[35,36], and leads to a BZF-QBICs mode exhibiting both polarization-independent optical behavior and high-Q factor[37]. By matching the energy of BZF-QBIC modes to that of *A* excitons via optimizing the thickness of the bulk $WS_2$ metasurfaces, typical anti-crossing features with the same Rabi splitting energies of 165.9 meV are observed in absorption spectra for arbitrarily polarized light, e.g., *x*- and *y*-linear polarizations, right and left circular polarizations, indicating that the BZF-QBIC-supported strong coupling holds polarization-independent characteristics. This work provides an efficient method for accessing polarization-insensitive strong coupling in TMDC materials, and paving the way for achieving TMDC-based chip-scale polaritonic devices with arbitrary polarizations.

## 2. Numerical Models

Figure 1(a) shows the self-hybridized system, patterned bulk $WS_2$ is placed onto a $SiO_2$ substrate. Bulk $WS_2$ is chosen for its high refractive index ($n = 4.4$ at $\lambda = 620$ nm) and large oscillator strength (compared with $MoS_2$, $MoSe_2$, $WSe_2$). The permittivity of bulk $WS_2$ can be described by the Lorentz oscillator model[19]

$$\varepsilon(E) = \varepsilon_B + \frac{f_0 E_{exc}^2}{E_{exc}^2 - E^2 - iE\Gamma_{exc}} \tag{1}$$

where $\varepsilon_B = 20$, $f_0 = 0.2$, $E_{exc} = 2$ eV, and $\Gamma_{exc} = 50$ meV represent the background permittivity, the oscillator strength, the resonant energy, and the full linewidth of $A$ excitons, respectively. The permittivity of the bulk $WS_2$ metasurface can be regarded as the superposition of the permittivity of a pure excitonic mode and that of optical mode supported by the background-index dielectric nanocavity. The optical mode contribution can be estimated by setting the oscillator strength $f_0 = 0$, while the pure excitonic mode contribution can be isolated by setting the background permittivity $\varepsilon_B = 1$. Strong coupling between these modes can be achieved by tuning the geometric parameters of the bulk $WS_2$ metasurfaces, such as the thickness lattice periods. More information about the permittivity of bulk $WS_2$ can be found in Figure S1. In our study, the optical properties of $WS_2$ metasurfaces are calculated by COMSOL MULTIPHYSICS based on finite element method (FEM). The periodic boundary conditions are applied along *x*- and *y*- directions, while perfectly matched layers are used along *z* direction.

# 3. Results and Discussion

## 3.1 Polarization-independent BZF-QBIC mode

To demonstrate the proposed bulk $WS_2$ metasurfaces supporting polarization-independent BZF-QBIC modes, the oscillator strength $f_0$ is set to zero (considering only the background refractive index). In the unperturbed case (Figure 1(a)), the metasurface consists of periodically arranged square unit cells with periods of $P_x = P_y = 165$ nm and a thickness of $H = 20$ nm, each unit cell contains a central circular hole with a radius $R = 50$ nm. These geometric parameters are chosen on purpose to achieve self-hybridized strong coupling between the optical mode and the $A$ excitonic resonance, as discussed below. When the perturbation parameter $d$ (the deviations of four adjacent holes from their original unit-cell center) is introduced (Figure 1(b)), the unit cell of the new structure consists of four circular holes. As a result, the periodicity of the unit cell at both $x$- and $y$- directions are doubled, meanwhile the first Brillouin zone of the metasurface is folded (as shown in the inset of Figure 1(c)). To demonstrate the BZF effect, the transverse-magnetic (TM) band structures of these cases are calculated. As shown in Figure 1(c), at the unperturbed conditions, the fundamental mode lies below the light line, indicating that this mode is guided mode (GM). When the BZF effect is introduced, this GM is folded above the light line. The QBIC feature of this folded mode can be confirmed by the electromagnetic field distribution at Γ point. As shown in Figure S2, the electric field exhibits even symmetry under $C_2$ operation, which possesses a symmetry incompatible with radiative states characterized by odd symmetry electromagnetic vectors under $C_2$ operation, indicating the QBIC signature of this folded mode. Therefore, such a folded mode is referred to as Brillouin zone folding induced QBIC mode. To understand the origins of this mode, a multipole decomposition and near-field distribution methods were employed for retrieving the fundamental electromagnetic contributions, i.e., electric dipole (ED, $P_\alpha$), magnetic dipole (MD, $M_\alpha$), electric toroidal dipole (ETD, $T_\alpha^{(e)}$), magnetic toroidal dipole (MTD, $T_\alpha^{(m)}$), electric quadrupole (EQ, $Q_{\alpha\beta}^{(e)}$), and magnetic quadrupole (MQ, $Q_{\alpha\beta}^{(m)}$) (Supplemental Material S3). As shown in Figure S3(a), the BZF-QBIC mode is dominated by ETD component, this also agree with the electromagnetic field distribution obtained by near-field analysis (Figure S3(b)). Compared to the ED mode, the electric field enhancement of the ETD mode is typically confined to the gaps between the holes, making ETD-dominated optical modes especially suitable for achieving strong light-matter coupling[38].

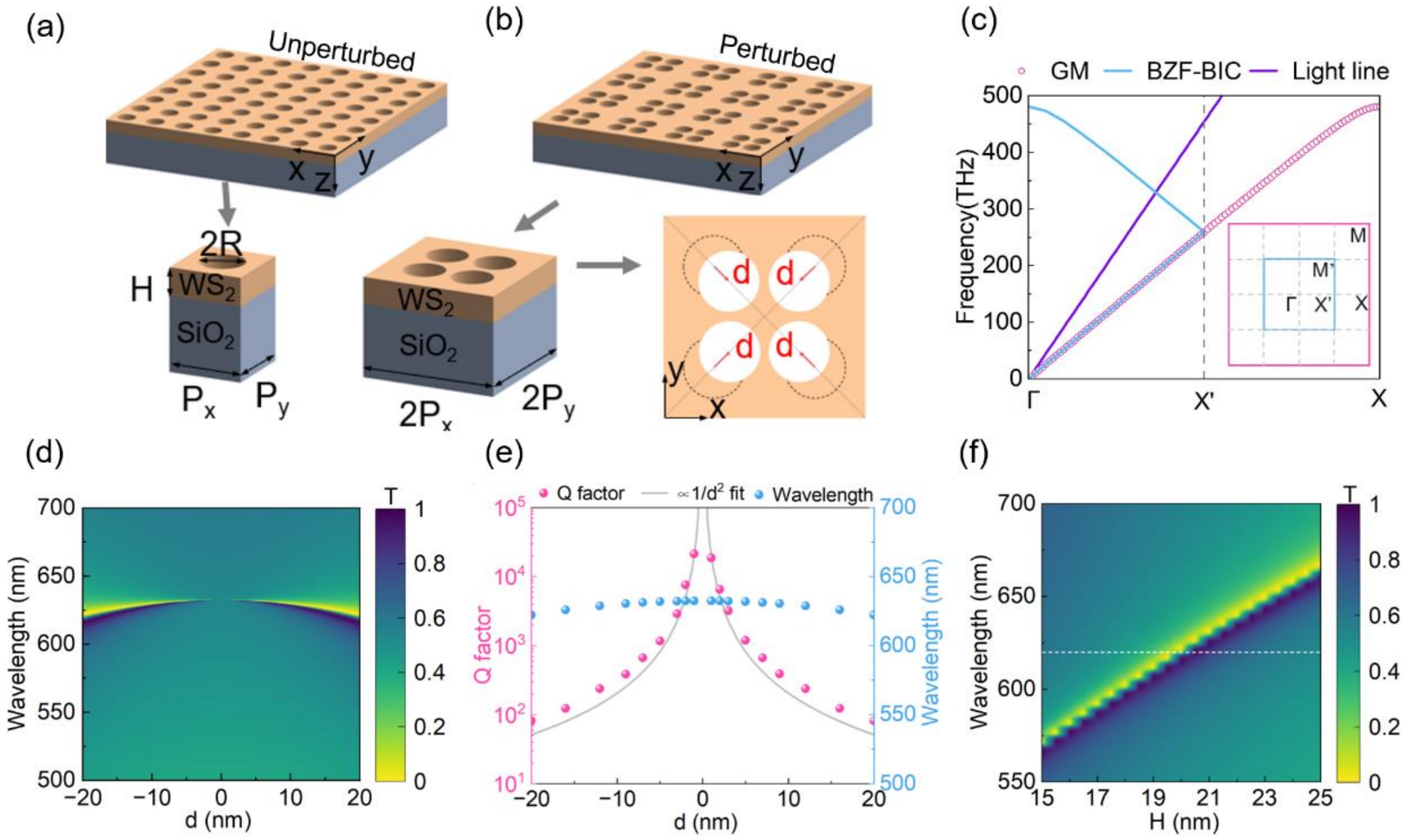

**Figure 1** Optical properties of the BZF-QBIC modes in the bulk $WS_2$ metasurfaces. Metasurfaces and corresponding unit cell: (a) Unperturbed condition; (b) Perturbed condition, where the perturbation parameter d is defined as the deviations of four adjacent holes from their original unit-cell center. (c) Transverse magnetic (TM) band structure of the unperturbed (red hollow circle) and perturbed (blue solid curve) cases, the inset shows first Brillouin zone corresponding to the unperturbed (red) and perturbed (blue) cases. (d) Absorption spectrum of the metasurfaces as a function of perturbation parameters $d$. (e) The Q-factors and wavelengths of the BZF-QBIC modes as a function of perturbation parameters $d$. (f) Absorption spectrum of the metasurfaces as a function of thickness $H$, while the white dashed lines represent the dispersion of excitons in bulk $WS_2$ (reference).

In order to maximize the self-hybridized coupling in bulk $WS_2$ metasurfaces, the influence of geometric parameters on wavelength and Q-factor of BZF-QBIC mode is investigated. As indicated in Figure 1(d), the optical mode of the bulk $WS_2$ metasurface shows a Fano-like lineshape (Figure S4) at the perturbed condition ($d \neq 0$ nm) in transmission spectrum of normally incident $x$-polarized plane waves, whereas it disappears at the unperturbed condition ($d = 0$ nm). Clearly, the linewidth of these modes increases with respect to $d$. To further understand the effect of perturbation parameter on BZF-QBIC modes, the Fano formula[39,40] is used to fit the transmission spectra to extract the Q factor of these modes (Figure S4). The wavelength of the BZF-QBIC modes keeps nearly unchanged when increasing the value of $d$, while the Q-factor of these modes shows an inverse square law with $d$ (Figure 1(e)), suggesting that an ideal high-Q optical cavity with a stable resonant wavelength for strong coupling can be obtained by adjusting the value of $d$. In addition, the dispersion behavior of the BZF-QBIC mode was also investigated by analyzing the transmission spectra of $WS_2$ metasurfaces with different thickness of $H$ (fix $d$). This mode redshifts when increasing $H$, whereas its linewidth remains nearly unchanged (Figure 1(f)). Furthermore, these geometric parameters ensure the spectral matching between the BZF-QBIC mode and *A* excitons of $WS_2$, providing a basis for achieving self-hybridized strong coupling in the following sections.

The above results are obtained by only considering the case of normally incident *x*-polarized light. To extend the results to arbitrary polarized optical illumination, as sketched in Figure 2(a), both linearly and circularly polarized illuminations are considered. In the case of linearly polarized irradiation, the polarization angle $\theta$ of the incident light is defined as the angle between the oscillation direction of the incident optical field and the *x*-axis. The transmission spectra of the metasurface are calculated with incident polarization angles ranging from 0° to 90°, with $d$ fixed at 20 nm (see Figures 2b and S5). The results clearly show that both resonant wavelength and Q-factor of the optical modes keep unchanged across the entire polarization angle. The same spectral features are also presented for the left-/right-handed circularly polarized (LCP/RCP) illumination (Figure 2(b)). Such polarization-insensitive transmission spectra of the metasurface confirm the polarization-independent property of the BZF-QBIC mode, which is related to $C_{4v}$ symmetry of the structure and the perturbation. This polarization independency can also be evidenced by analyzing the constituents of BZF-QBIC mode from a multipole decomposition and the near field distribution. As shown in Figure 2(c), the ETD component dominates the BZF-QBIC mode irrespective of the polarization state of incident light (*x*-, *y*-, LCP and RCP). Moreover, the near-field distributions are unchanged for different polarized irradiations (Figure 2(d)). Especially for LCP and RCP illumination, the same local field distribution can be understood from the fact that both LCP and RCP light consists of equally *x*- and *y*-polarized components but with a phase difference of π. This point towards the existence of the modes by incident with more general elliptically polarized light, that is superposition of two orthogonally linearly polarized components with a fixed phase difference. The polarization-independent property of the BZF-QBIC mode is especially critical to realize strong coupling with arbitrary polarizations.

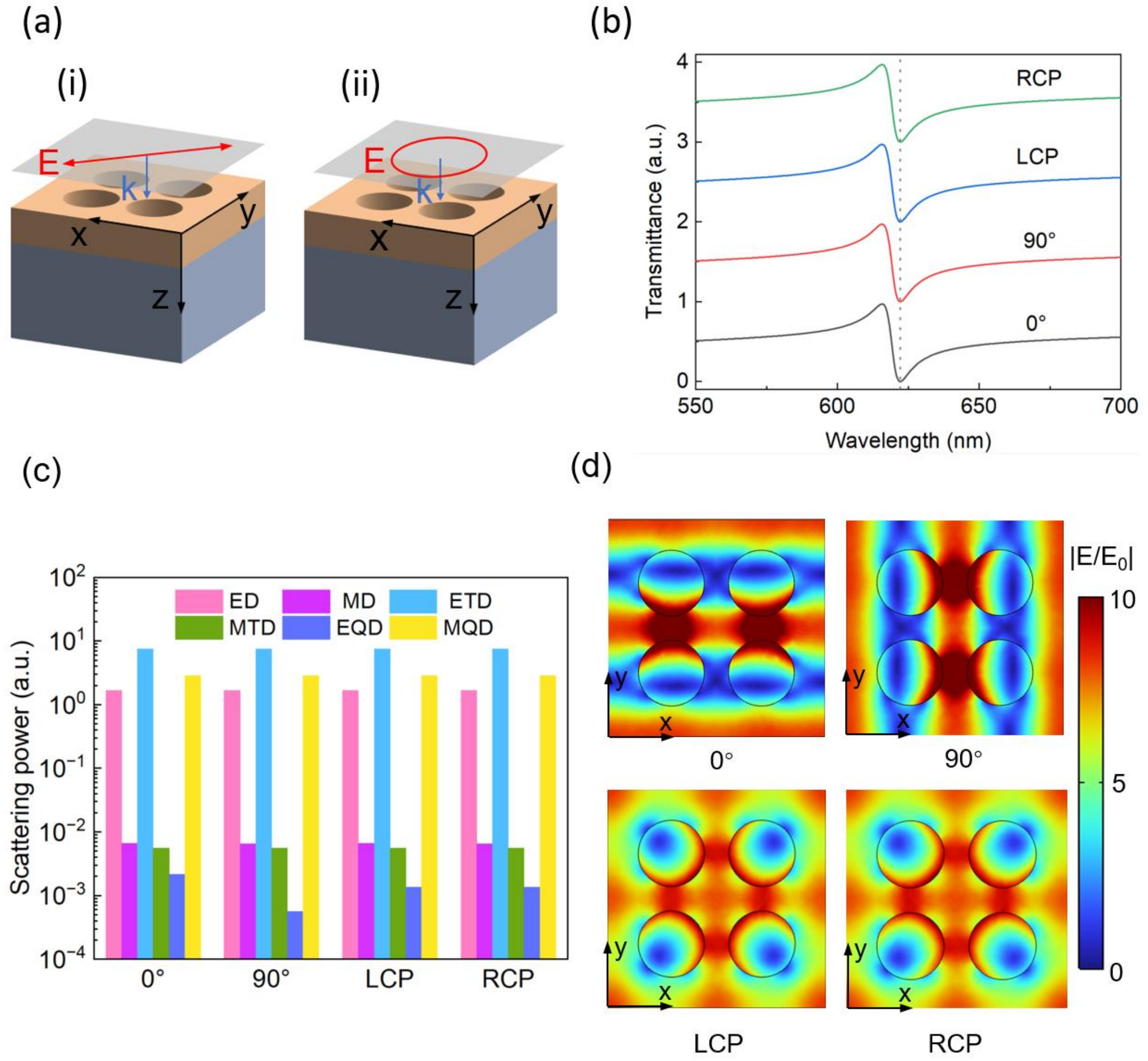

**Figure 2** Polarization-independent BZF-QBIC mode. (a) Schematic diagrams of the metasurface illuminated at normal-incidence with (i) linearly polarized and (ii) circularly polarized light. (b) The transmission spectra of the metasurface under *x*-, *y*-polarization, left-/right-hand circular polarization (LCP/RCP) illumination. (c) Multipole decomposition and (d) Near-field electric field distribution of BZF-QBIC modes under the same four incident polarization states.

## 3.2 Polarization-independent self-hybridized strong coupling

To explore the polarization-independent strong coupling in the self-hybridized bulk $WS_2$ metasurfaces, the refractive index of both the background-index dielectric nanocavity that supports optical modes and the pure excitonic mode are taken into account ( $f_0 = 0.2$ ). The geometric parameters of these metasurfaces ( $P = 330\,\text{nm}$, $d = 20\,\text{nm}$, $R = 50\,\text{nm}$) are intentionally chosen to facilitate the spectral overlap between $A$ excitons and BZF-QBICs mode. The absorption spectra of these bulk $WS_2$ metasurfaces are calculated as a function of $H$ under $x$-polarized illumination, as shown in Figure 3(a). These spectra exhibit two distinct anticrossing branches with respect to excitonic mode, indicating a signature of strong coupling between $A$ excitons and the BZF-QBIC mode. The peak of these branches with wavelengths smaller/greater than the excitonic wavelength correspond to upper/lower polaritons (UP/LP). Notably, a residual peak between UP and LP arises from the fraction of uncoupled $A$ excitons.

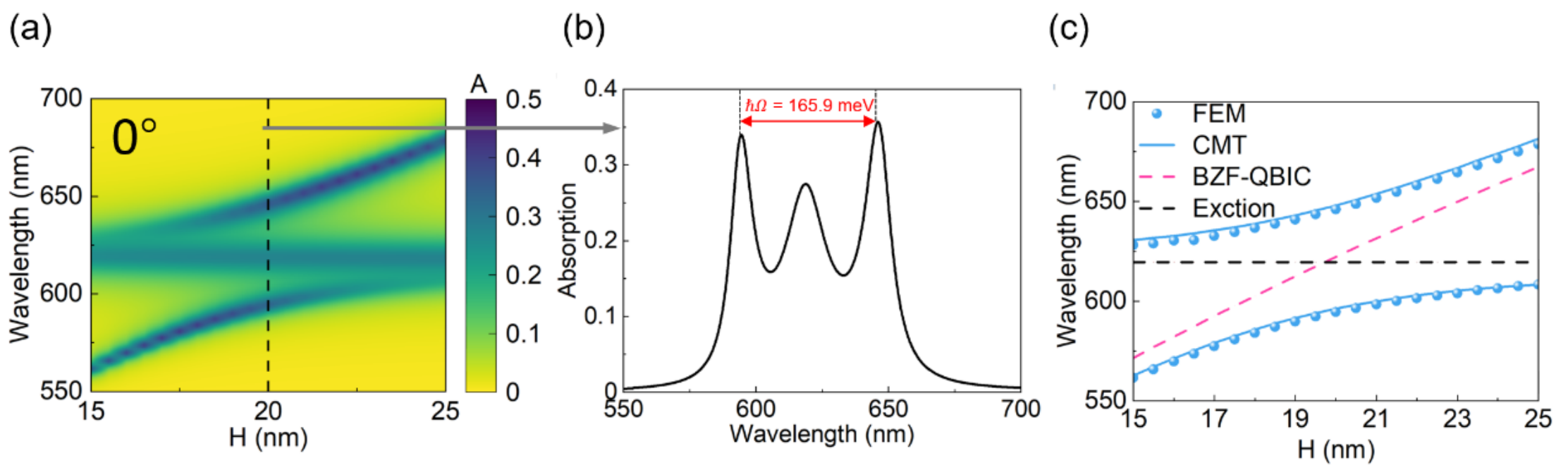

**Figure 3** Optical properties of the self-hybridized strong coupling. (a) Absorption spectra of the self-hybridized coupling as a function of thickness $H$ of the metasurface under normally incident $x$-polarized light (0°). The black dashed line corresponds to (b) the absorption spectrum of the self-hybridized structure at zero detuning. (c) Dispersion relations of the self-hybridized coupling system obtained from FEM calculation (blue dots) and CMT model (blue solid curves). The black and red dashed lines represent the dispersion relations of the uncoupled $A$ excitons and BZF-QBIC mode, respectively.

To confirm the self-hybridized strong coupling in bulk $WS_2$ metasurfaces, the coupled oscillator model (CMT) is uesd[41]

$$\begin{pmatrix} E_{cav} + i\hbar\gamma_{cav} & g \\ g & E_{exc} + i\hbar\gamma_{exc} \end{pmatrix} \begin{pmatrix} \alpha \\ \beta \end{pmatrix} = E_{UP,LP} \begin{pmatrix} \alpha \\ \beta \end{pmatrix} \quad (2)$$

here, the energies and half-widths at half-maximum (HWHM) of the uncoupled BZF-QBIC mode and the *A* excitons are denoted by $E_{cav}$, $E_{exc}$ and $\gamma_{cav}$, $\gamma_{exc}$, respectively. The coupling strength between the BZF-QBIC mode and the excitons is represented by $g$. Moreover, $\alpha$ and $\beta$ are the Hopfield coefficients satisfying the normalization condition $|\alpha|^2 + |\beta|^2 = 1$, where $|\alpha|^2$ and $|\beta|^2$ represent the relative contributions of the light and matter components to the polaritons, respectively. When the coupled system satisfies zero detuning $E_{cav} = E_{exc}$, the Rabi splitting can be obtained as:

$$\hbar\Omega = E_{UP} - E_{LP} = 2\sqrt{g^2 - \frac{1}{4}(\hbar\gamma_{cav} - \hbar\gamma_{exc})^2} \quad (4)$$

The following conditions must be satisfied for strongly coupled system:

$$c_1 = \frac{g}{\frac{|\gamma_{cav} - \gamma_{exc}|}{2}} > 1 \quad (5)$$

$$c_2 = \frac{g}{\sqrt{\frac{\hbar\gamma_{cav}^2 + \hbar\gamma_{exc}^2}{2}}} > 1 \quad (6)$$

The former inequation ensures the existence of the Rabi splitting, while the latter one guarantees its observability in experiments. A Rabi splitting of 165.9 meV can be extracted from the results of zero detuning, as shown in Figure 3(b). Moreover, the coupling strength of the system is estimated to be 83.2 meV when the dissipation of both the BZF-QBIC mode and the *A* excitons is taken into account. Therefore, the designed bulk $WS_2$ metasurface satisfies the strong coupling criteria (Inequations (5) and (6)). In addition, the existence of strong coupling in this metasurface is further evidenced by the agreement between FEM calculations and the CMT model (Figure 3(c)). Benefiting from the ETD-dominated high-Q BZF-QBIC mode, the proposed self-hybridized strong coupling system exhibits a large Rabi splitting and a stronge coupling strength[32], enabling a broader operating wavelength range for polaritonic devices.

In order to investigate the polarization-independent strong coupling for the bulk

$WS_2$ metasurface, the illuminated light is extended from $x$-polarization to other linearly polarized light with arbitrary $\theta$, and even to circularly polarized light (LCP and RCP). As shown in Figures 4(a-c), the absorption spectra of the bulk $WS_2$ metasurfaces are calculated as a function of $H$ under $y$-polarized, LCP, and RCP conditions. The spectra show the same anticrossings and Rabi splittings for all polarization states, indicating the same strongly coupled characteristics under these polarizations. To further study the strong coupling behavior under various polarized light illumination, the absorption spectra of the metasurface irradiated with different polarized light, i.e., linear polarization with various $\theta$ and LCP and RCP, are compared at zero detuning, as displayed in Figure 4(d). A constant of Rabi splitting of 165.9 meV is observed under all polarized conditions. Considering the polarization-independent behavior of both BZF-QBIC mode and excitonic mode, the coupling strength of the hybrid system remains a constant of 83.2 meV. In addition, the linewidth of each polaritonic branch is also the same at different polarizations. These characteristics confirm the polarization-independent property of strong coupling in bulk $WS_2$ metasurface.

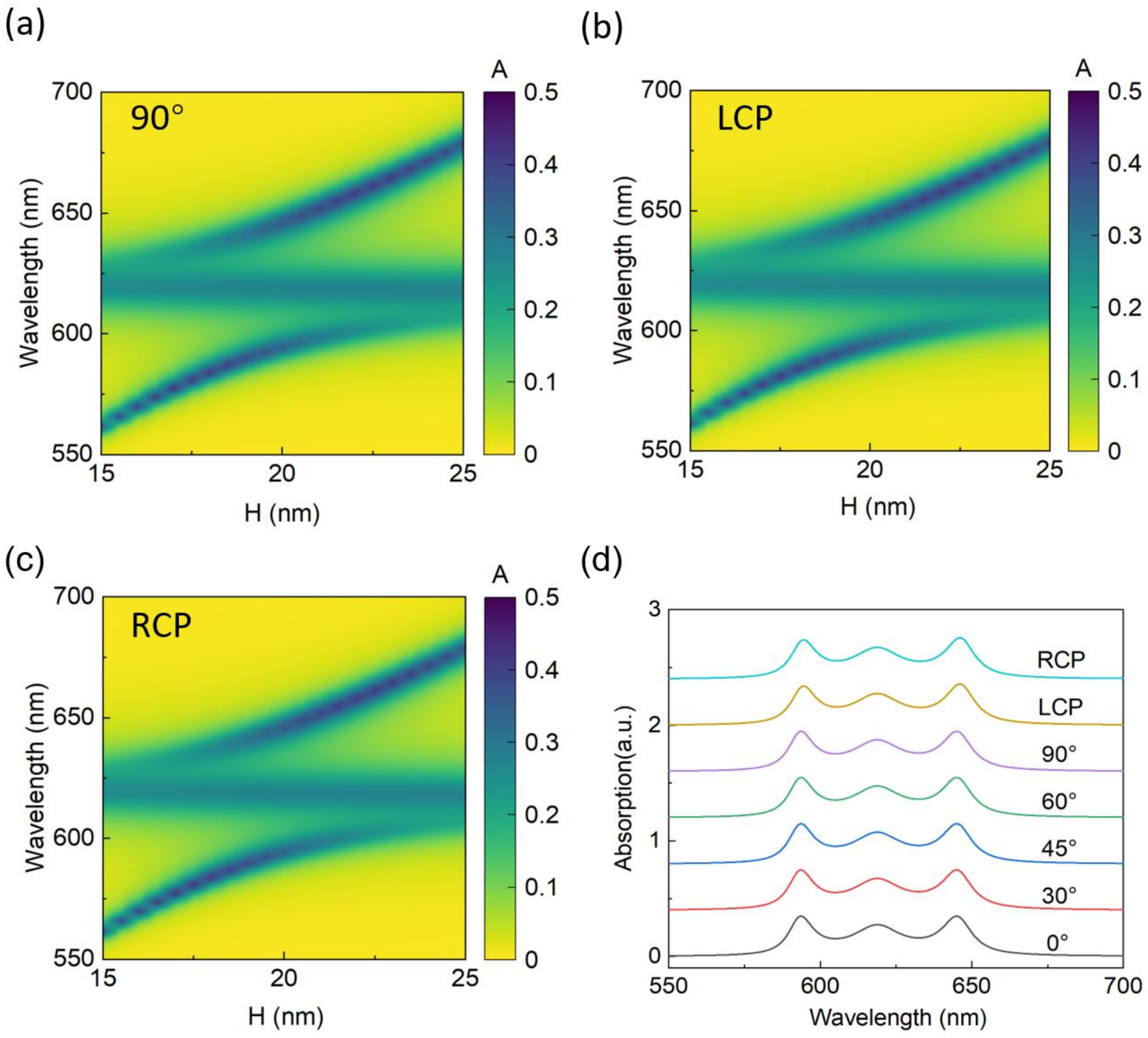

**Figure 4** Polarization-independent self-hybridized strong coupling in bulk $WS_2$ metasurfaces. (a) Absorption spectra of the self-hybridized structure as a function of the thickness H of the metasurface under normal incidence with (a) 90°($y$-polarized), (b) LCP and (c) RCP. (d) Absorption spectra of the coupled system at zero detuning illuminated with linearly polarized, LCP, and RCP light.

The polarization-insensitive strong coupling in this study provides a significant advance over previously reported TMDC systems[28-32], in which the maximum coupling strength can only be achieved with incident light at a specific polarization owing to the polarization-sensitive Mie resonances. Furthermore, the proposed polarization-independent BZF-QBIC originates from Brillouin-zone folding rather than the intrinsic properties of materials, is not limited to the self-hybridized bulk $WS_2$ platform and can be extended to other frequency regimes for application in other excitonic or vibrational systems. Recently, a $C_{4v}$-symmetric $WSe_2$ metasurface supporting degenerate QBIC mode and enabling polarization-independent strong coupling has been reported[42]. Their QBIC mode is dominate by toroidal resonance, which is different with the ETD-dominated BZF-QBIC in our case. Moreover, their mode is quite sensitive to the structural asymmetry parameter, leading to a clear spectral shift and strong detuning of the strong coupling by precisely changing the asymmetry, thereby hindering the realization of robust polaritonic devices.

In this study, a perturbation parameter insensitive, polarization-independent self-hybridized strong coupling is achieved in bulk $WS_2$ metasurface. However, such polarization- independent strong coupling can only be excited under normal incidence. Designing a structure that can simultaneously support angle-independent along with polarization-independent BZF-QBIC modes, thereby achieving angle- and polarization-independent strong coupling in two-dimensional materials, will further expand the application scenarios of polaritonic devices. This is possible because the excitonic modes in TMDCs are nondispersive and polarization-independent.

## 4. Conclusions

In summary, we propose a novel polarization-independent self-hybridized strong coupling system between *A* excitons and BZF-QBIC mode in the bulk $WS_2$ metasurface. This polarization-independent, high-Q BZF-QBIC mode with an asymmetry-insensitive resonant wavelength, supported by the proposed metasurface, is particularly suitable for self-hybridized strong coupling. Our results also demonstrate that self-hybridized strong coupling is achieved with a Rabi splitting of 165.9 meV in bulk $WS_2$ metasurface. Furthermore, the self-hybridized strong coupling in this metasurface is confirmed to be polarization-independent under excitation with different polarization states. Our work provides a new method for realizing polarization-independent strong coupling in bulk TMDCs metasurfaces, offering new insights for expanding the applications of low-dimensional polaritonic devices with arbitrary polarizations.

## 5. Acknowledgements

We acknowledge support from the National Natural Science Foundation of China under Grant No. 12304371 and the Young Talent Support Plan of Xi'an Jiaotong University.

Supplemental Material

# Polarization-independent self-hybridized strong coupling by Brillouin zone folding induced quasi-bound states in the continuum in bulk $WS_2$ metasurfaces

Xiongwei Guo, Fu Liu, Jinhai Si, Xun Hou, Kuidong Wang*

*Key Laboratory for Physical Electronics and Devices of the Ministry of Education and Shaanxi Key Lab of Information Photonic Technique, School of Electronic Science and Engineering, Xi'an Jiaotong University, Xi'an, 710049, China*

*Contact author: wangkuidong@xjtu.edu.cn

## S1. Dispersive permittivity of bulk $WS_2$.

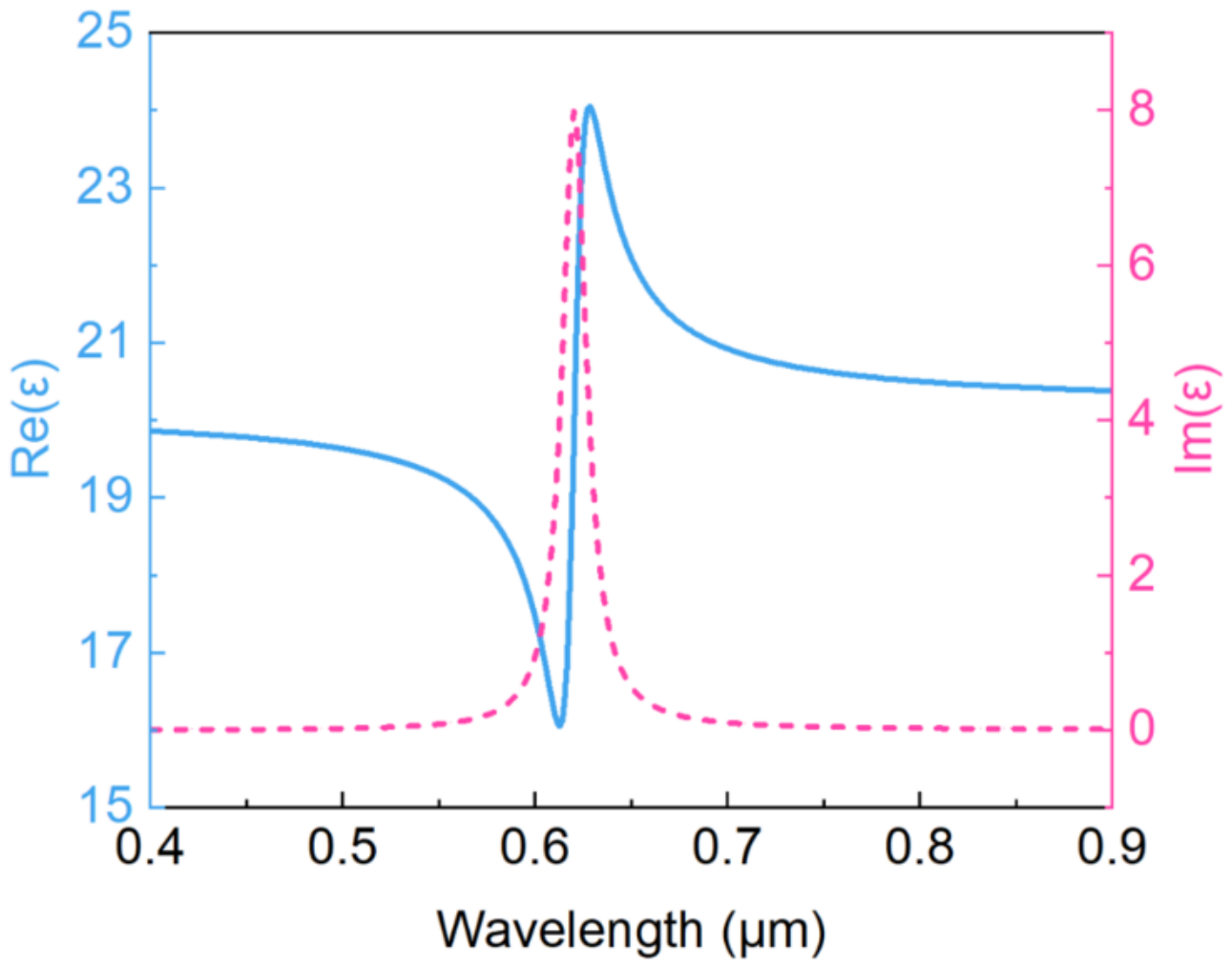

**Figure S1.** The real (blue solid curve) and imaginary (red dashed curve) parts of the permittivity for bulk $WS_2$.

## S2. Electric field distribution at Γ point under Brillouin zone folding condition.

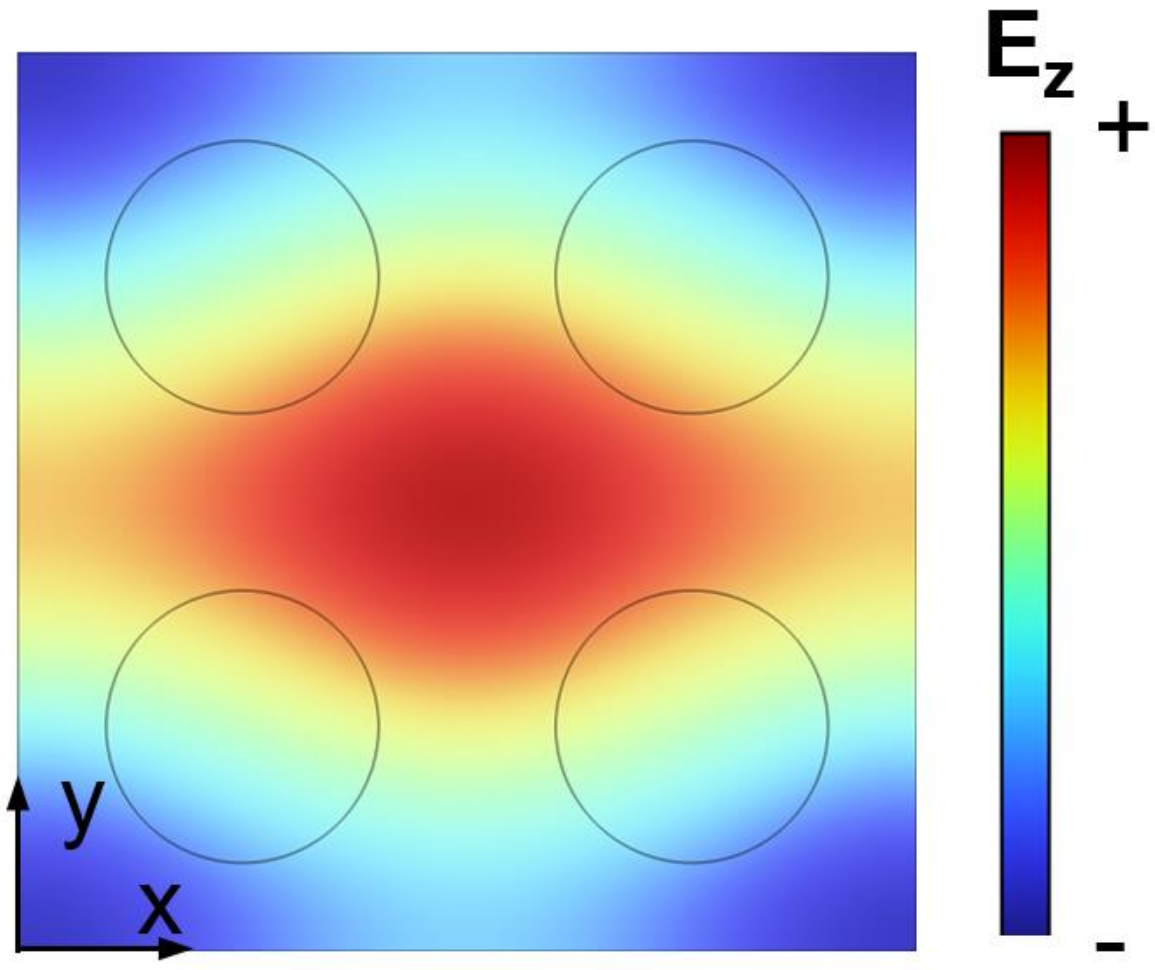

**Figure S2.** Electric field distribution at Γ point under Brillouin zone folding condition.

## S3. Multipole decomposition and electromagnetic field distribution of the BZF-QBIC mode.

We analyzed the far-field radiation of the BZF-QBIC mode by the standard multipole decomposition method, and calculated the contributions from the electric dipole (ED, $P_\alpha$), magnetic dipole (MD, $M_\alpha$), electric toroidal dipole (ETD, $T_\alpha^{(e)}$), magnetic toroidal dipole (MTD, $T_\alpha^{(m)}$), electric quadrupole (EQ, $Q_{\alpha\beta}^{(e)}$), and magnetic quadrupole (MQ, $Q_{\alpha\beta}^{(m)}$). In Cartesian coordinate system ($\alpha, \beta = x, y, z$), ED, MD, ETD, MTD, EQ, and MQ can be expressed as

$$P_\alpha = \frac{1}{i\omega}\int \vec{j} d^3r \quad (1)$$

$$M_\alpha = \frac{1}{2c}\int (\vec{r}\times\vec{j}) d^3r \quad (2)$$

$$T_\alpha^{(e)} = \frac{1}{10c}\int [(\vec{r}\cdot\vec{j})\vec{r} - 2r^2\vec{j}] d^3r \quad (3)$$

$$T_\alpha^{(m)} = \frac{k^2}{20}\int r^2(\vec{r}\times\vec{j}) d^3r \quad (4)$$

$$Q_{\alpha\beta}^{(e)} = \frac{1}{2i\omega}\int \left[r_\alpha j_\beta + r_\beta j_\alpha - \frac{2}{3}(\vec{r}\cdot\vec{j})\delta_{\alpha,\beta}\right] d^3r \quad (5)$$

$$Q_{\alpha\beta}^{(m)} = \frac{1}{3c}\int \left[(\vec{r}\times\vec{j})_\alpha r_\beta + (\vec{r}\times\vec{j})_\beta r_\alpha\right] d^3r \quad (6)$$

$c, r, \omega, n, k, \vec{E}$ and $\vec{j} = i\omega(n^2 - 1)\vec{E}$ denote the speed of light, the location, the angular frequency of light, the refractive index, the wave vector, the electric field, and the current density, respectively. The scattering powers corresponding to each multipole moment contribution can be given as[1,2]

$$I_P = \frac{2\omega^4}{3c^3}|P_\alpha|^2 \tag{7}$$

$$I_M = \frac{2\omega^4}{3c^3}|M_\alpha|^2 \tag{8}$$

$$I_{T^{(e)}} = \frac{2\omega^6}{3c^5}|T_\alpha^{(e)}|^2 \tag{9}$$

$$I_{T^{(m)}} = \frac{2\omega^6}{3c^5}\left|T_\alpha^{(m)}\right|^2 \tag{10}$$

$$I_{Q^{(e)}} = \frac{\omega^6}{5c^5}\sum|Q_{\alpha\beta}^{(e)}|^2 \tag{11}$$

$$I_{Q^{(m)}} = \frac{\omega^6}{40c^5}\sum|Q_{\alpha\beta}^{(m)}|^2 \tag{12}$$

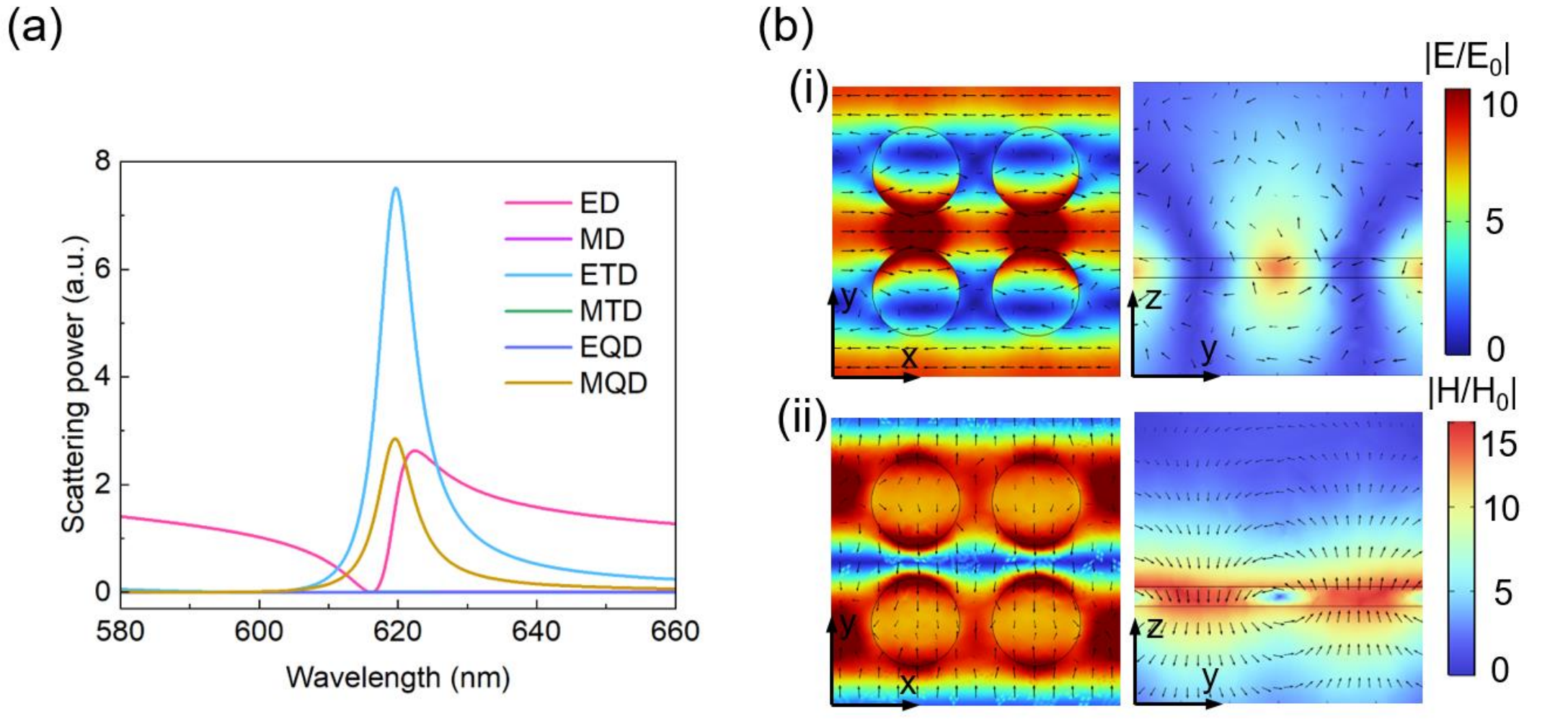

**Figure S3.** (a) Multipolar decomposition of the BZF-QBIC mode. (b) Distribution of (i) electric field and (ii) magnetic field of the BZF-QBIC mode.

## S4. Fitting transmission spectrum using Fano formula

When fix the perturbation parameter, e.g., $d = 20$ nm, the simulated transmission spectrum can be fitted (gray solid curve) by the Fano formula

$$T_{\text{Fano}}(\omega) = \left|a_1 + ja_2 + \frac{b}{\omega - \omega_0 + j\gamma}\right|^2 \tag{1}$$

where $a_1$, $a_2$ and $b$ are constant real numbers, $\omega_0$ is the central resonant frequency. The radiation dissipation $\gamma$ can be obtained from the Fano line shape, from which the Q factor of optical mode in bulk $WS_2$ metasurfaces can be determined.

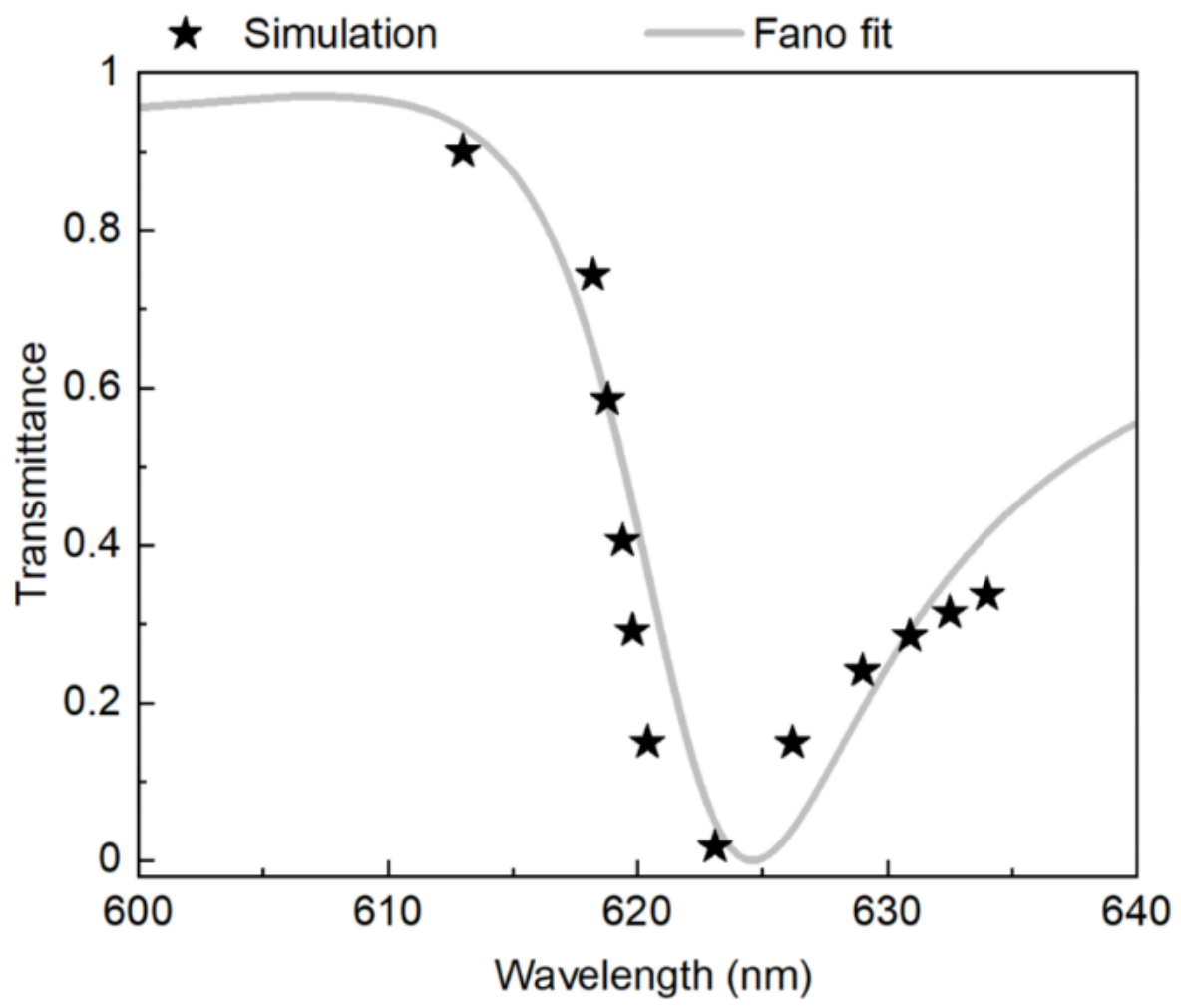

**Figure S4.** Transmission spectrum of the bulk $WS_2$ metasurface simulated and fitted using Fano formula for a perturbation parameter of 20 nm.

## S5. Transmission spectra as a function of polarization angle

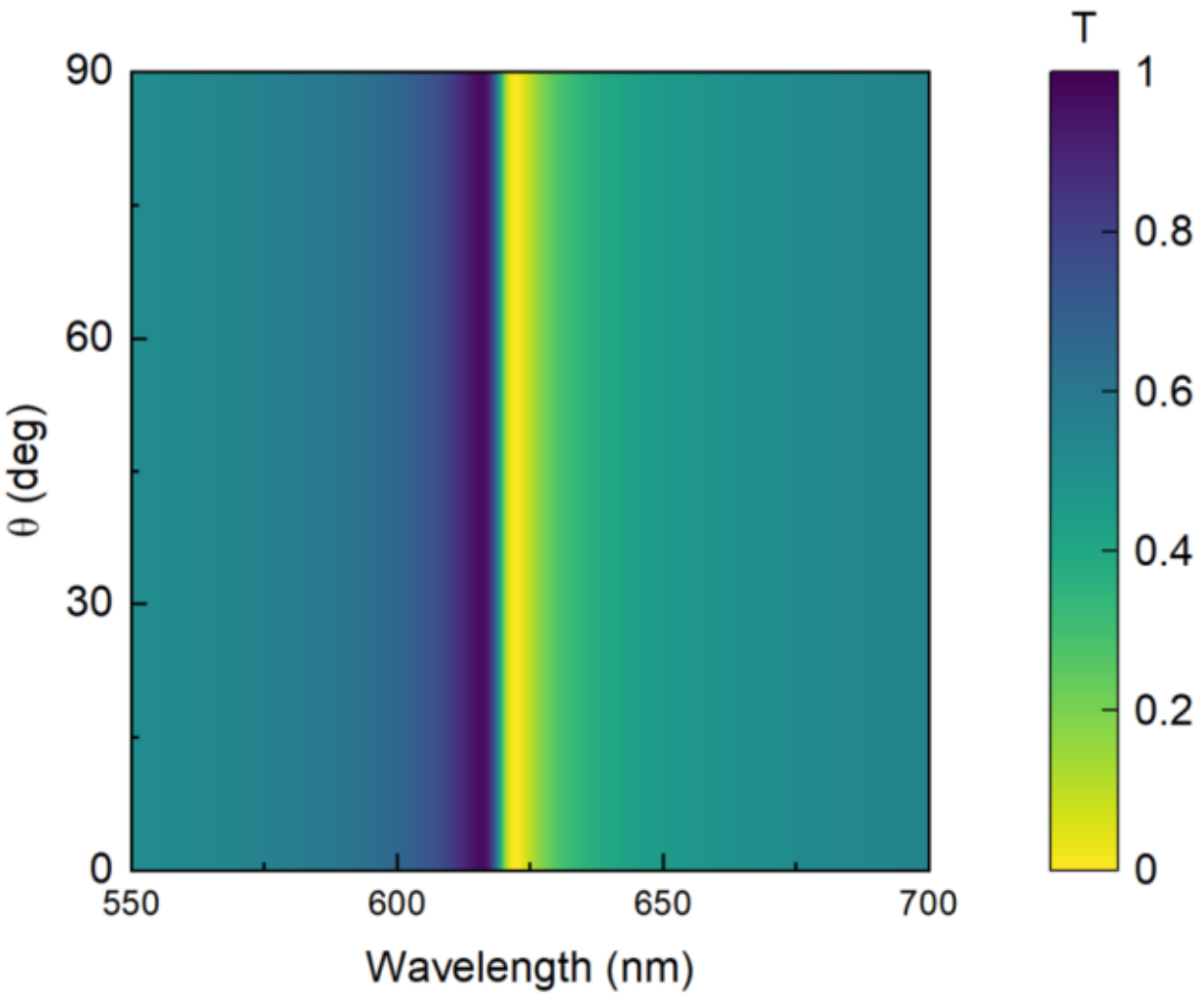

**Figure S5.** Transmission spectra as a function of the polarization angle $\theta$ of the linearly polarized incident light.